\documentclass[preprint,12pt]{article}

\usepackage{amssymb}
\usepackage{amsmath}
\usepackage{moreverb,url}
\usepackage{natbib}
\setcitestyle{square,numbers}
\usepackage[colorlinks,bookmarksopen,bookmarksnumbered,citecolor=red,urlcolor=red]{hyperref}
\usepackage{siunitx}
\usepackage[ruled,vlined,linesnumbered]{algorithm2e}
\usepackage{tikz} %Flow charts
\usepackage{soul}
\usepackage{soulutf8}
\usetikzlibrary{shapes,positioning, arrows, backgrounds, fit}
\usepackage{authblk} %Add authors and affiliations

\usepackage{listings} % Programming Listings
\usepackage{xcolor}
\definecolor{codegreen}{rgb}{0,0.6,0}
\definecolor{codegray}{rgb}{0.5,0.5,0.5}
\definecolor{codepurple}{rgb}{0.58,0,0.82}
\definecolor{backcolour}{rgb}{0.95,0.95,0.92}
\lstdefinestyle{mystyle}{ %Style of programming listings
	language=C++,
	backgroundcolor=\color{backcolour},   
	commentstyle=\color{codegreen},
	keywordstyle=\color{magenta},
	numberstyle=\tiny\color{codegray},
	stringstyle=\color{codepurple},
	basicstyle=\ttfamily\footnotesize,
	breakatwhitespace=false,         
	breaklines=true,                 
	captionpos=b,                    
	keepspaces=true,                 
	numbers=left,                    
	numbersep=5pt,                  
	showspaces=false,                
	showstringspaces=false,
	showtabs=false,                  
	tabsize=2
}
\newcommand{\vecb}[1]{\mathbf{#1}}
\newcommand{\vvec}{\vecb{v}}
\newcommand{\xvec}{\vecb{x}}
\newcommand{\phasespacevar}{\xvec,\vvec}

\title{A performance portable implementation of the semi-Lagrangian algorithm in six dimensions}

\author[1]{Nils Schild}
\author[1]{Mario Räth}
\author[2]{Sebastian Eibl}
\author[1]{Klaus Hallatschek}
\author[3]{Katharina Kormann}
\affil[1]{Max Planck Institut for Plasma Physics, Germany}
\affil[2]{Max Planck Computing and Data Facility, Germany}
\affil[3]{Ruhr University Bochum, Germany}
\date{} 
\begin{document}

	\maketitle
	
	\begin{abstract}
		In this paper, we describe our approach to develop a simulation software application for the fully kinetic Vlasov equation which will be used to explore physics beyond the gyrokinetic model. Simulating the fully kinetic Vlasov equation requires efficient utilization of compute and storage capabilities due to the high dimensionality of the problem. In addition, the implementation needs to be extensibility regarding the physical model and flexible regarding the hardware for production runs. We start on the algorithmic background to simulate the 6-D Vlasov equation using a semi-Lagrangian algorithm. The performance portable software stack, which enables production runs on pure CPU as well as AMD or Nvidia GPU accelerated nodes, is presented. The extensibility of our implementation is guaranteed through the described software architecture of the main kernel, which achieves a memory bandwidth of almost 500~GB/s on a V100 Nvidia GPU and around 100~GB/s on an Intel Xeon Gold CPU using a single code base. We provide performance data on multiple node level architectures discussing utilized and further available hardware capabilities. Finally, the network communication bottleneck of 6-D grid based algorithms is quantified. A verification of physics beyond gyrokinetic theory for the example of ion Bernstein waves concludes the work.
	\end{abstract}
		
	\section{Introduction}
	
	In this paper, we discuss our effort in providing a platform-independent and flexible semi-Lagrangian solver for the fully kinetic Vlasov equation in six-dimensional phase space for strongly magnetized plasmas. With the simulation software we investigate plasma turbulence phenomena, which are not captured by commonly used models, as in the article of \citet{Raeth2023}. Since common models for magnetized plasma (such as occur in magnetic Fusion) reduce the dimensionality of the Vlasov equation by assuming strong restrictions on either the temporal or spatial
	scales, interesting physics phenomena might be missed. Gyrokinetic theory, a widely used model in this area, averages over the gyroradius removing phenomena at the Larmor frequency. Upcoming large scale supercomputers offer the capabilities to store and compute simulations based on the full Vlasov equation capturing phenomena on small spatial scales with high frequencies. Our goal is to develop a plasma simulation application, which allows us to investigate phenomena beyond gyrokinetics in high frequency regimes.
	
	We build the implementation based on the \texttt{Kokkos} framework developed by \citet{Trott2022}, since this framework has a long term team of developers and a large user base. This way the code can profit from the adaption of the \texttt{Kokkos} kernels to future hardware and we do not have to develop and maintain a performance portability layer. We follow the optimization techniques in \citet{Kormann2019} to utilize hardware performance capabilities with our implementation. Next to the performance portability results we discuss the software architecture of the main kernel using design patterns. The idea of design patterns originates from \citet{Gamma2007}. Through the usage of design patterns we implement a modularized kernel which leads to an extensible as well as testable implementation. We verify the implementation by reproducing the dispersion relation of ion Bernstein waves which are waves with the magnitude of the Larmor frequency.
	% and more recent work can be found in \citet{Iglberger2022}
	
	\subsection{Relation to previous work}
	
	Recently multiple small and large software projects have used GPUs to accelerate numerical simulations. Several approaches, which are closely or loosely related our approach, have been taken to write performance portable code. 
	
	Established software frameworks which can not be ported through a full rewrite implement their own performance portability layer to support their main data structures or use directive based approaches to execute their kernels on GPU. GENE \citep{Germaschewski2021} and AMReX \citep{AMReX2019} would be examples for the first approach while the latter porting technique is used by ORB5 \citep{ORB52021}. Our approach is related to the Cabana Toolkit \citep{Cabana2023} or the Alpine miniapps \citep{Alpine2022} for particle applications. Both create datastructures based on the performance portability framework \texttt{Kokkos} \citep{Trott2022} to use multiple shared memory techniques. Alternatives to \texttt{Kokkos} would be Raja \citep{Raja2019}, which has been investigated in \citep{Artigues2020}, or Alpaka \citep{Alpaka2017}.
	
	High dimensional implementations of semi-Lagrangian algorithms have been presented in the articles of \citet{Bigot2013}, \citet{Umeda2014} and \citet{Kormann2019} for the 5-D and 6-D Vlasov equation. Both codes focused on CPU scaling. A performance portable miniapp based on a semi-Lagrangian algorithm of the GYSELA code has been ported by \citet{Yuuichi2019}. \citet{Einkemmer2020} has implemented a semi-Lagrangian algorithm for 4-D problems based on the discontinuous Galerkin method using OpenMP for CPUs and CUDA for GPUs. Our work extends the shared memory concepts by the HIP programming model leading to a comparison of OpenMP, CUDA and HIP backends of \texttt{Kokkos}. Scaling results of the implementation of \citet{Einkemmer2020} from 4-D to 6-D problems are given by \citet{Einkemmer2022}. We encounter similar difficulties when scaling 6-D semi-Lagrangian algorithms to multiple nodes and quantify the communication bottleneck. 
	
	\subsection{Outline of the paper}
	The outline of the reminder of this paper is as follows: In the next section, we introduce the problem and the algorithmic background. Section \ref{sec:software_stack_parallelization} discusses the software stack and required background information as well as usage of \texttt{Kokkos} in the BSL6D implementation. Moreover, we review the challenges of a domain decomposition in six dimensions. Our software architecture of the main kernel, a distributed Lagrangian interpolation, based on design patterns is  presented in section \ref{sec:software_architecture}. The performance of our novel code is analyzed on various platforms concerning both its shared and distributed memory performance in sections \ref{sec:shared_performance} and \ref{sec:distributed_performance}, respectively. The scaling results are quantitatively validated with the network communication bottleneck. We demonstrate that the code can provide physical results beyond the gyrokinetic model in section \ref{sec:physical_validation}. The results show in particular that high order interpolation stencils in the semi-Lagrangian method are of importance in practical applications. Finally, section \ref{sec:conclusions} summarizes the conclusions from the reported experiments.
	
	\section{Problem formulation and algorithmic background}

\subsection{The Vlasov equation}

A kinetic description of the motion of a plasma is given by the Vlasov--Maxwell system in the book of \citet[p.115ff]{LandauLifshitz1981}: The plasma is described by a distribution function $f_s$ in phase space for each species of charge $q_s$ and mass $m_s$ which evolves in self-consistent and external electromagnetic fields. The self-consistent fields evolve according to Maxwell's equations. In this paper, we focus on an electro-static description neglecting the self-consistent magnetic field. %To simplify the numerics we start with the electro static Vlasov-Poisson equation coupled to a constant background magnetic field $\vecb{B}_0$
Moreover, we consider the distribution functions for electrons in an inert
neutralizing ion background. Using rationalized electrostatic CGS units, then the electron distribution evolves according to the Vlasov equation
\begin{align}
\label{equ:vlasov}
&\partial_t f(\phasespacevar,t) + \vvec \cdot \nabla_\xvec f(\phasespacevar,t) \nonumber \\
&+ \dfrac{q}{m}(\vecb{E}(\xvec,t) + \vvec \times \vecb{B}_0) \cdot \nabla_\vvec f(\phasespacevar,t) = 0, \text{ on } \Omega \times \mathbb{R}^3 \times (0,T)
\end{align}
where $\Omega$ is the spatial domain, $\vecb{B}_0$ denotes the constant
background magnetic field and $\vecb{E}(\xvec,t)$ denotes the electric field
based on the Poisson equation for the electric potential
$\phi$,
%The Vlasov-Poisson equation can be used to describe a single particle species (e.g. ions, electrons). In the following we consider electrons and therefore set $q$ to $-1$. The Vlasov equation is then coupled to the Poisson equation
\begin{align}
\rho(\xvec,t) &= q\left(\int f(\phasespacevar,t) \mathrm{d}\vvec - n_0\right) \nonumber\\
\label{equ:poisson}
- \Delta \phi(\xvec,t) &= \rho(\xvec,t) \\
\vecb{E}(\xvec,t) &= -\nabla\phi(\xvec,t) \nonumber.
\end{align}
The right-hand-side of the Poisson equation is given by the charge density
$\rho$ and $n_0$ is the density of oppositely charged particles in the
neutralizing background. 

The characteristic curves of the Vlasov equation \eqref{equ:vlasov} are the solutions of the ordinary differential equations
\begin{align}
\label{equ:characteristic_odes}
\dfrac{\mathrm{d}\vecb{X}}{\mathrm{d}t} = \vecb{V} && \dfrac{\mathrm{d}\vecb{V}}{\mathrm{d}t} = \dfrac{q}{m}\left(\vecb{E}\left(\vecb{X},t\right) + \vecb{V} \times \vecb{B}_0 \right).
\end{align}
As a hyperbolic conservation law, the Vlasov equation~\eqref{equ:vlasov} conserves the particle distribution function $f$ along the characteristic curves in phase space. Solving \eqref{equ:characteristic_odes} with initial conditions $(\phasespacevar)$ at time $t$, we denote the resulting characteristic curves by $(\vecb{X}(s;\phasespacevar,t),\vecb{V}(s;\phasespacevar,t))$. We can follow these curves backward in time to an initial condition $f_0$ at time 0. Using the conservation properties along the characteristic curves gives a mapping of the values of $f$ at $t$ to the values of $f_0$ at time $0$. Inverting this mapping allows us to get an expression of the solution of equation~\eqref{equ:vlasov} at time $t$ as a function of the initial conditions $f_0$ at time $0$. 
\begin{align}\label{equ:f_from_f0}
f(\phasespacevar,t) = f_0(\vecb{X}(0;\phasespacevar,t),\vecb{V}(0;\phasespacevar,t)).
\end{align}
Since the characteristic curves are depending on the electric field which in turn is depending on the distribution function, this expression cannot be used for practical calculations of the solution.
%Hereby $f(\phasespacevar,t)$ with $(\phasespacevar)\in \mathbb{R}^{3\times3}$ denotes the particle distribution function in 6-D phase space, $\rho$ the electron charge density. We consider an neutralizing ion background (!!!Quelle ion neutralizing background rhoion=1!!!) and therefore set the ion charge density to $1$. The electrostatic potential is $\phi$ and electric field $\vecb{E}$. The constant background magnetic field $\vecb{B}_0$ is either zero or aligned with the $x_3$ axis.\\

\subsection{The split-step backward semi-Lagrangian method}
\label{sec:sl}
The idea of the semi-Lagrangian method is to discretize the phase space and use the conservation properties of \eqref{equ:vlasov} to propagate the distribution function on the grid from $t_m$ to $t_{m+1} = t_m + \Delta t$. We denote the grid points by $(\xvec_i,\vvec_j)$. Applying the semi-Lagrangian method consists of the following two steps.

First the characteristic equations~\eqref{equ:characteristic_odes} are solved backward in time from $t_{m+1}$ to $t_m$ using a point $(\vecb{x}_i,\vvec_j)$ as an initial condition. Secondly, the obtained characteristic curve is plugged into the right-hand side of mapping~\eqref{equ:f_from_f0} to update the value of $f$ at the grid point $(\xvec_i,\vvec_j)$ at time $t_{m+1}$ using the value of $f$ at time $t_m$ at the foot of the characteristic
\begin{align}
f(\vecb{x}_i,\vvec_j, t_{m+1}) = f(\vecb{X}(t_m;\xvec_i,\vvec_j,t_{m+1}),\vecb{V}(t_m;\xvec_i,\vvec_j,t_{m+1}),t_m).
\end{align}
Usually  $f(\vecb{X}(t_m;\xvec_i,\vvec_j,t_{m+1}),\vecb{V}(t_m;\xvec_i,\vvec_j,t_{m+1}),t_m)$ is not located at a grid point at time $t_m$. Therefore, this value has to be interpolated using the values $\{f(\xvec_i,\vvec_j,t_m)\}$ located at the grid points at time $t_m$.

The semi-Lagrangian method as described up to now requires solving equation~\eqref{equ:characteristic_odes} and a 6-D interpolation. Instead, $\xvec$ and $\vvec$ advection operators can be separated using splitting methods as presented in the articles of \citet{Cheng1976, McLachlan2002}. We use a Strang splitting, which has second order accuracy in time, to split $\xvec$ and $\vvec$ advection. Separating $\xvec$ and $\vvec$ advection has the advantage that the electric field---necessary to solve the characteristic equation of the velocity step---remains unchanged in the velocity advection step. Thus the characteristic equations of the subsystems can be solved analytically. Furthermore, the remaining 3-D interpolation steps contain only commuting operators. The commuting operators can be further decomposed into three 1-D interpolations through a Lie splitting.
%Solving the ODE directly is problematic since $\vecb{E}(\xvec,t)$ depends on $f$ through equation~\eqref{equ:poisson} and a 6-D interpolation is computationally expensive.\citep[p.70ff]{Sonnendruecker2021}(!!!Bessere Quelle Semi Lagrange!!!)\\

%\subsection{Poisson equation}
The Vlasov--Poisson system also involves solving the Poisson equation~\eqref{equ:poisson} to determine the electric field $\vecb{E}$. The computational effort to solve this three dimensional problem is negligible compared to the effort needed for the advection equation. So far only periodic boundary conditions have been implemented. The Poisson problem is solved by a pseudo-spectral method based on the Fast Fourier Transform.

%Several interpolation methods can be used for the interpolation step. We focus on Lagrange interpolation here, since 

\subsection{Lagrange interpolation}
\label{subsec:Lagrange_Interpolation}

Several interpolation schemes can be used for the 1-D interpolations. We follow the article of \citet{Kormann2019} and use Lagrange interpolation due to its locality and accuracy. Spline interpolation is often preferred in simulations of the Vlasov equations due to its increased smoothness. However, this interpolation is global and also in the localized form discussed by \citet{Kormann2019}, the increased communication costs make spline interpolation less competitive in large scale simulations. Another alternative would be discontinuous interpolations described by \citet{crouseilles2011} which, need numerous points before the become competitive due to their decreased smoothness. 

Let us now consider the Lagrange interpolation in 1-D and let $x_j$, $j=1,\ldots,N$, be the grid points, equidistantly spaced with distance $\Delta x$, and $\alpha$ some displacement for $x_j$. We want to compute the interpolant $f(x_j+ \alpha)$ from values of $f$ surrounding $f(x_j)$ assuming $\vert \alpha \vert< \Delta x$. The interpolant can be compared to a local stencil algorithms. Furthermore, we denote by $\ell_i^q$ the Lagrange-polynomials of order $(q-1)$ with $q$ nodes in the interpolant. We distinguish two cases:
\begin{itemize}
	\item For an odd number $q$, the interpolation is given by
	\begin{align}
	\label{equ:lagrange_interpolation_odd}
	f(x_j+\alpha) = \sum_{i=j-(q-1)/2}^{j+(q-1)/2} \ell_i(\alpha) f(x_i)
	\end{align}
	\item For an even number $q$, we consider an interpolation stencil centered around the interpolated point $x_j+\alpha$, such that the interpolation is given by:
	\begin{align}
	\label{equ:lagrange_interpolation_even}
	f(x_j + \alpha) = 
    	\begin{cases}
              \sum_{i=j-q/2}^{j+q/2-1} l_i(\alpha )  f(x_i) & \alpha \leq 0 \\
              \sum_{i=j-q/2+1}^{j+q/2} l_i(\alpha )  f(x_i) & \alpha > 0 \\
    	\end{cases}
	\end{align}
\end{itemize}
We note that the second stencil is centered around the foot of the characteristic and is therefore an upwinding-type scheme which generally gives better results. Higher order stencils lead to better resolution of physical phenomena which is shown in figure~\ref{fig:ion_bernstein_wave_dispersion} of section~\ref{sec:physical_validation}. At the same time they increase computational complexity and the communication overhead.

The interpolation shift $\alpha$ obtained from a characteristic as discussed in section~\ref{sec:sl} is a function of a lower-dimensional subset $D\subset\mathbb{R}^6$, where $\mathrm{dim}(D)\in\{1,2,3\}$, and the corresponding mapping $\alpha: \text{D} \rightarrow \mathbb{R}$. The most general way to define the interpolation shift is as a function of the full phasespace $\mathbb{R}^6$ (and not just $D$). Using the mapping $\alpha: \mathbb{R}^6 \rightarrow \mathbb{R}, \alpha = \alpha(\phasespacevar )$ eases abstractions in the software architecture.
In subsection~\ref{subsec:software_arch_lag_intrp} a software architecture will be presented to implement the 1-D interpolation using multiple interpolation shifts $\alpha $ and stencils with different width $q$.

\subsection{Characteristics of the algorithmic steps}
\label{subsec:bsl6d_algorithm}
\begin{figure}[!htb]
\begin{center}
\resizebox{\linewidth}{0.5\linewidth}{
\begin{tikzpicture}
[auto,
 block/.style ={rectangle, thick, text width=10em, align=center, rounded corners, minimum height=2em},
 mergeblock/.style ={rectangle, draw=black, thick, text width=10em, align=center, rounded corners, minimum height=2em, inner sep=2ex},
 textblock/.style ={rectangle, text width=10em, align=center},
 line/.style ={draw, thick, -latex',shorten >=2pt}]
 \node [block, minimum width=0.3\textwidth, fill=blue!20, draw=blue] 
        (halox) {Halo communication in $x_i$\\
        \scriptsize{MPI: Point to Point}};
 \node [textblock, above=of halox.north, anchor=south] 
 		(loopx) {Loop over $i=\{ 1,2,3 \}$};
 \node [block, minimum width=0.3\textwidth, below=of halox.south, anchor=north, fill=green!20, draw=green] 
 		(intrpx) {Interpolate $f(\phasespacevar )$ in $x_i$};
 \begin{scope}[on background layer]
	\node[mergeblock, fit=(loopx) (halox) (intrpx), inner xsep=1ex]
	      (xspace) {};
 \end{scope}

 \node [block, minimum width=0.3\textwidth, right=of halox.east, anchor=west, fill=red!20, draw=red] 
 		(charge) {$\rho = \int f(\phasespacevar ) \mathbb{d}^3v$\\
 		\scriptsize{MPI: Allreduce (on $v$-topology)}};
 \node [textblock, above=of charge.north, anchor=south] 
(timestep) {Propagate from $t$ to $t+\Delta t$};
 \node [block, minimum width=0.3\textwidth, right=of intrpx.east, anchor=west, fill=brown!20, draw=brown] 
  		(field) {Solve field equations\\
  		\scriptsize{MPI: All to All (on $x$-topology)}};

 \node [block, minimum width=0.3\textwidth, right=of charge.east, anchor=west, fill=blue!20, draw=blue] 
        (halov) {Halo communication in $v_i$\\
        \scriptsize{MPI: Point to Point}};
 \node [textblock, above=of halov.north, anchor=south] 
 		(loopv) {Loop over $i=\{ 1,2,3 \}$};
 \node [block, minimum width=0.3\textwidth, below=of halov.south, fill=green!20, draw=green] 
 		(intrpv) {Interpolate $f(\phasespacevar )$ in $v_i$};
 \begin{scope}[on background layer]
	\node[mergeblock, fit=(loopv) (halov) (intrpv), inner xsep=1ex] 
	      (vspace) {};
 \end{scope}

 \node[mergeblock, fit=(xspace) (charge) (timestep) (field) (vspace), inner xsep=1ex] 
	       (prop) {};

 \draw (halox) edge[->,very thick] (intrpx);
 \path (intrpx.east) edge[->,thick] (charge.west);
 \path (charge) edge[->,very thick] (field);
 \path (field.east) edge[->,thick] (halov.west);
 \path (halov) edge[->,very thick] (intrpv);
\end{tikzpicture}
}
\end{center}
\caption{\label{fig:flowchart} Algorithmic steps needed to propagate the distribution function by a single time step $\Delta t$. The color coding highlights the algorithmic steps which are observed separately in subsection~\ref{subsec:weak_scaling}.}
\end{figure}
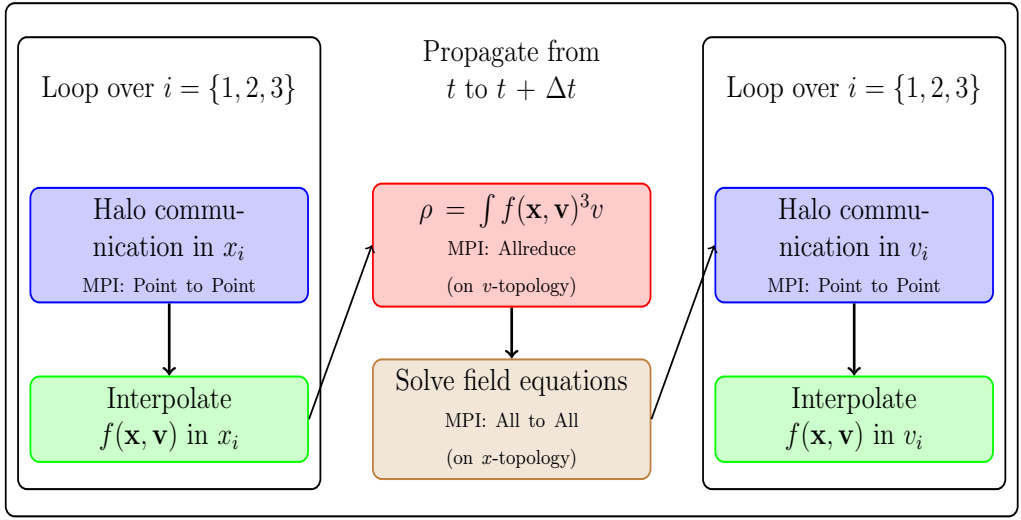

Before discussing the implementation of the above discussed algorithm in our novel BSL6D code, let us summarize the algorithmic steps as they are compiled together in a setting where the 6-D domain is decomposed into blocks that are distributed with the help of MPI. figure~\ref{fig:flowchart} summarizes the steps. Concerning computations, we distinguish steps on the 6-D data (distribution function), steps on the 3-D data (fields) and mappings between both of them. For the communication, we distinguish point-to-point, All-to-All and Allreduce, while the latter two are taking place on subgroups of the 6-D MPI distribution.

%\begin{figure}
%\begin{center}
%\includegraphics[scale=0.5]{./Pictures/FlussdiagrammBSL6D.png}
%\caption{\label{fig:flowchart} Algorithmic steps needed to propagate the distribution function by a single time step $\delta t$}
%\end{center}
%\end{figure}
	\section{Software stack and parallelization}
\label{sec:software_stack_parallelization}

\subsection{Software stack}
The development of a scalable application for high dimensional problems involves shared and distributed parallel concepts. In addition, care has to be taken about input and output of the application. A reliable application has to be tested intensively. These challenges are met through several third party libraries on which the BSL6D code relies. In the following the necessary libraries will be introduced.

An important feature of the code will be a node level independent implementation. Production runs of the BSL6D algorithm containing CPU only as well as GPU accelerated nodes. Different shared memory concepts like \texttt{OpenMP}, \texttt{CUDA} or \texttt{HIP} are needed to access these architectures. The performance portability framework \texttt{Kokkos} \citep{Trott2022} is used to abstract the node level architecture from the main implementation.

\texttt{Kokkos} is a library based approach to offer performance portability with \texttt{C++}. How the BSL6D implementation makes use of \texttt{Kokkos} will be described in the next subsection. So far the BSL6D code has been successfully tested on different shared memory architectures listed in table~\ref{tab:systems_under_consideration}. 

\begin{figure}
	\begin{center}
		\begin{tikzpicture}[auto, node distance=0.0cm,
			block/.style={rectangle, draw, text centered, 
				rounded corners, minimum height=2em, inner xsep=2pt},
			concept/.style={rectangle, draw, text centered, minimum width=0.075\textwidth,
				rounded corners, minimum height=1em, inner xsep=2pt}]
			\node [block, fill=blue!80, minimum width=\textwidth] (bsl6d) {BSL6D};
			\node [block, minimum width=0.25\textwidth, below=of bsl6d.south west, 
			fill=blue!50, anchor=north west] (kokkos) {Kokkos};
			\node [block, minimum width=0.15\textwidth, right=of kokkos.east, 
			fill=blue!50, anchor=west] (mpi) {MPI};
			\node [block, minimum width=0.15\textwidth, right=of mpi.east, 
			fill=blue!25, anchor=west] (hdf5) {HDF5};
			\node [block, minimum width=0.25\textwidth, right=of hdf5.east, 
			fill=blue!25, anchor=west] (heffte) {HeFFTe};
			\node [block, minimum width=0.2\textwidth, right=of heffte.east, 
			fill=blue!10, anchor=west] (gtest) {GoogleTest};
			\node [concept, below=of kokkos.south west, 
			fill=red!20, anchor=north west] (cuda) {\scriptsize{CUDA}};
			\node [concept, right=of cuda.east, 
			fill=brown!20, anchor=west] (omp) {\scriptsize{OpenMP}};
			\node [concept, right=of omp.east, 
			fill=green!20, anchor=west] (hip) {\scriptsize{HIP}};
			\node [concept, below=of heffte.south west, 
			fill=red!20, anchor=north west] (cufft) {\scriptsize{cufft}};
			\node [concept, right=of cufft.east, 
			fill=brown!20, anchor=west] (fftw) {\scriptsize{fftw}};
			\node [concept, right=of fftw.east, 
			fill=green!20, anchor=west] (rocfft) {\scriptsize{rocfft}};
		\end{tikzpicture}
		\caption{\label{fig:softwareStack}Software Stack of the BSL6D Code including the tested backends for regarding on performance portability}
	\end{center}
\end{figure}
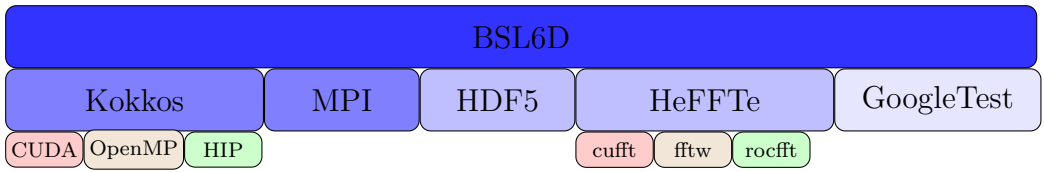

While \texttt{Kokkos} provides an abstraction for shared memory paradigms scaling to large computing clusters requires distributed memory concepts for which we rely on the widely used and well defined MPI standard \citep{MPIthecompletereference1998}. The input and output is based on the HDF5 library \citep{hdf5}. An FFT library is needed to solve the Poisson equation~\eqref{equ:poisson}. For this we use the HeFFTe library \citep{Aylan2020} which provides an interface to different vendor specific FFT libraries like cuFFT, rocFFT and FFTW. In addition HeFFTe adds distributed transforms to the previously mentioned solvers which allows us to completely offload distributed Fourier Transforms to HeFFTe. The full software stack is given in figure~\ref{fig:softwareStack}.

To ensure flexibility and reliability of the application we make use of the GoogleTest \texttt{C++} library and a test driven development approach.

\subsection{\texttt{Kokkos} performance portability framework}
As described in the previous subsection the implementation is performance portable across multiple node level architectures or shared memory concepts through the usage of \texttt{Kokkos} \citep{Trott2021}. In the following section we briefly recap the features of \texttt{Kokkos} needed within this paper. Detailed explanations are given in the \texttt{Kokkos} Documentation \citep{KokkosDocs,Trott2021}.

The memory management in the BSL6D implementation relies on \verb`View`s introduced by \texttt{Kokkos}, which behave like a multidimensional array with shared ownership first introduced by \citet{Edwards2014}. The data stored within a \verb`View` resides in a \verb`MemorySpace`. Using memory spaces \texttt{Kokkos} abstracts different memory resources like high bandwidth memory (HBM) of GPUs, DRAM of a CPUs or memory concepts like \texttt{Cuda Unified Virtual Memory} (UVM). At compile time \texttt{Kokkos} sets a default memory space in which data of a \verb`View` resides if no explicit memory space is defined. This default memory space depends on the backend \texttt{Kokkos} has been compiled for. If \texttt{Kokkos} has been compiled with the \texttt{CUDA} backend data within \verb`View`s by default resides on the Nvidia GPU. If \texttt{Kokkos} has been compiled with the \texttt{OpenMP} backend data resides in CPU memory.

An important memory space used in this paper is the \verb`ScratchSpace` which is shared memory on GPUs. This manageable memory allows to explicitly prefetch data into cache memory and therefore enables to optimize memory access for e.g. transpositions with discontinuous memory access as described by \citet{MatrixTranspose2022}. The scratch space can be accessed within a \verb`TeamPolicy` which is one of the \verb`ExecutionPolicies` provided by \texttt{Kokkos}. The execution policy defines how to iterate through an index range in a parallel kernel. The \verb`TeamPolicy` allows for nested levels of parallelism like vectorization on CPUs or usage of \texttt{CUDA} blocks which enables to tune block sizes to enable architecture specific optimization techniques at runtime.

A similar concept to memory spaces for \verb`View`s exists for execution policies with \verb`ExecutionSpaces`. The execution space defines where threads execute work. The execution space of an execution policy is defined through a \texttt{C++} template argument. \texttt{Kokkos} provides default execution spaces if no template argument is specified. In the BSL6D implementation kernels are executed in the default execution space. Therefore, not only computationally expensive but all kernels are accelerated. In addition, compiling \texttt{Kokkos} with a GPU compatible backend reduces memory copies between CPU and GPU, since \verb`View`s which are accessed by a kernel can also be kept in GPU memory throughout the computation.

The last point we have to take into account when implementing the BSL6D algorithm using \texttt{Kokkos} is input and output to hard drive. Care has to be taken about data exchange between different runs of the BSL6D algorithm and post processing of simulation data. \verb`View`s map indexes to memory addresses using \verb`MemoryLayouts`. The memory layout defines which index resides continuous in memory and which index has strided access. The memory layouts depend on the parallel backend \texttt{Kokkos} has been compiled for. Therefore, it is not fixed throughout independent runs of the BSL6D implementation, and we have to decide how to work with data stored on hard drive. We either need to include information how \verb`View`s have been stored or we fix the layout of data stored on hard disk. From a users perspective it is easier to work with a fixed layout than taking care about varying layouts. The layouts of multidimensional arrays defined in Mathematica and NumPy, which are the main post-processing tools we use, both use the C-Layout by default for newly created arrays\citep{MathematicaArray,NumpyArray}. We, therefore, stick to this convention and store our data on disk using the C-Layout. This corresponds to the \verb`LayoutRight` defined by \texttt{Kokkos} in which the right most index resides continuous in memory. Depending on the Kokkos backend a transposition step can be necessary to write the data correctly to disk.

\subsection{Domain decomposition for 6-D domains}
\label{subsec:domainDecomposition_6D}
Production runs on a 6-D phase space grid have huge memory requirements. Therefore, it is necessary to distribute the grid over multiple nodes on a compute cluster if high resolution runs are required. The distributed memory parallelism is provided through MPI. The BSL6D algorithm is based on the Cartesian Topology\citep[p.~319]{MPIthecompletereference1998} of the MPI standard. Load balancing is ensured by restricting the number of points per dimension to be chosen as a multiple of the number of processes in the corresponding dimension of the Cartesian Topology.

\begin{figure}
\begin{center}
\includegraphics[scale=0.5]{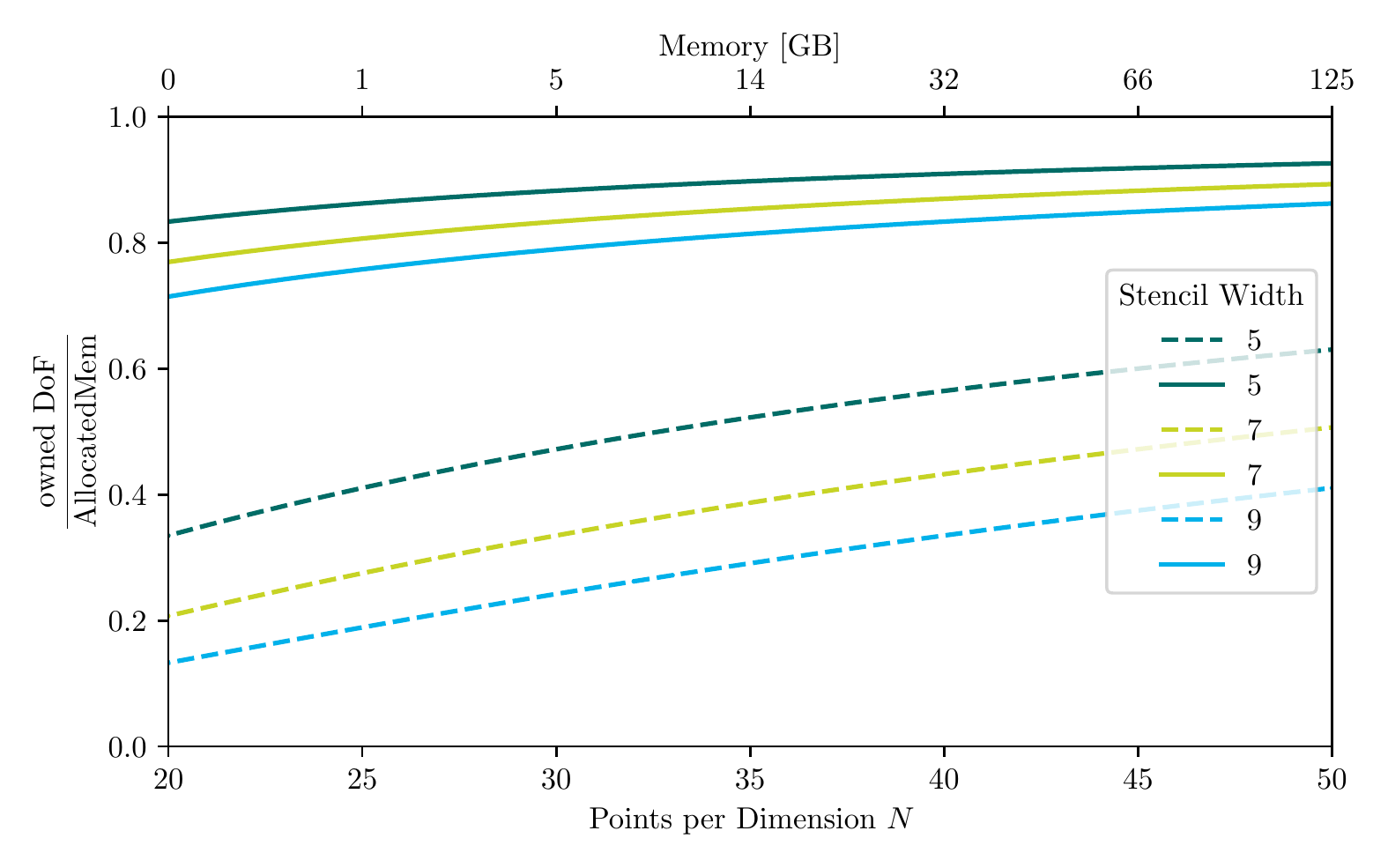}
\caption{\label{fig:memoryRatio}Ratio of owned degrees of freedom to memory requirements plotted against points per dimension $N$ on abscissa. For the dashed line the ratio is $N^6/(N+2w)^6$ without memory requirement optimization and for the solid lines it is $N^6/(N^6+2N^5w)$ which contains the memory optimization of subsection~\ref{subsec:domainDecomposition_6D}. The second abscissa gives the storage requirements for the owned degrees to show the high memory requirements of a 6D phasespace grid.}
\end{center}
\end{figure}

On the distributed grid the Lagrange interpolant needs to be evaluated. No MPI process owns the full phase space grid. We will refer to the data which is local to a processor as computation domain. Evaluating the stencil at the boundary of the computation domain requires grid points which are located at another processor. Therefore, independent processes have to exchange grid points. We refer to these regions as halo cells. Two options are available to implement halo cells. Either the \verb`View` containing the data of the computation domain could be extended by the halo regions or a new \verb`View` can be allocated to hold the data needed for the halo regions as suggested by \citet{Kormann2019}. Including the halo cells into the buffer which holds the distribution function would be the straight forward way used e.g. by \citet{Hager2011,Datta2009}. The disadvantage of this approach is the memory requirement in the high dimensional 6-D case. Assuming a 6-D hypercube with $N$ points in every dimension and halo width $w$ in every direction the allocated View would contain $(N+2w)^6$ points on each MPI-Rank. The ratio of computation domain to the size of the View is given in figure~\ref{fig:memoryRatio} through the dashed lines.

If the allocated memory is separated into one \verb`View` for the computation domain and one \verb`View` for the halo regions, two memory optimizations are available. On one hand no corners are allocated for the 6-D halo domain. In case of 1-D interpolations the corners in the halo domain are never used during for the interpolation. So the corners of halo regions would be wasted if they were allocated in the current implementation. On the other hand only halo regions in one of the six dimensions are needed during the 1-D interpolation. Therefore, it is sufficient to only allocate memory for the largest halo domain and share allocated memory between halo regions for interpolations in different dimensions. In case of the 6-D hypercube with $N$ grid points in every dimension with the boundary width $w$ the  total size of allocated \verb`View`s sums up to $N^6+2wN^5$. The ratio size of computation domain to size of allocated \verb`View`s is plotted in figure~\ref{fig:memoryRatio} with the straight line and the efficiency of memory usage has increased significantly.

A drawback of this separation is an enhanced complexity of the implementation. The interpolation stencil should iterate through continuous memory, but the halo and computation domain do not reside in contiguous memory any longer. Therefore, the memory optimization prohibits that the iteration is carried out using a single 1-D loop as would be possible with a stencil on contiguous memory as in the work of \citet{Hager2011,Datta2009}. A solution to this is discussed in the next section together with the software architecture for the interpolation algorithm.
	\section{Software architecture of the interpolation kernel}\label{sec:software_architecture}

\subsection{Performance considerations}
\label{subsec:performance_considerations}
Before the software architecture of the main kernel, the 1-D Lagrange interpolation, is discussed two performance critical components of the algorithm are examined.

\begin{figure}
	\begin{center}
		\includegraphics[scale=0.4]{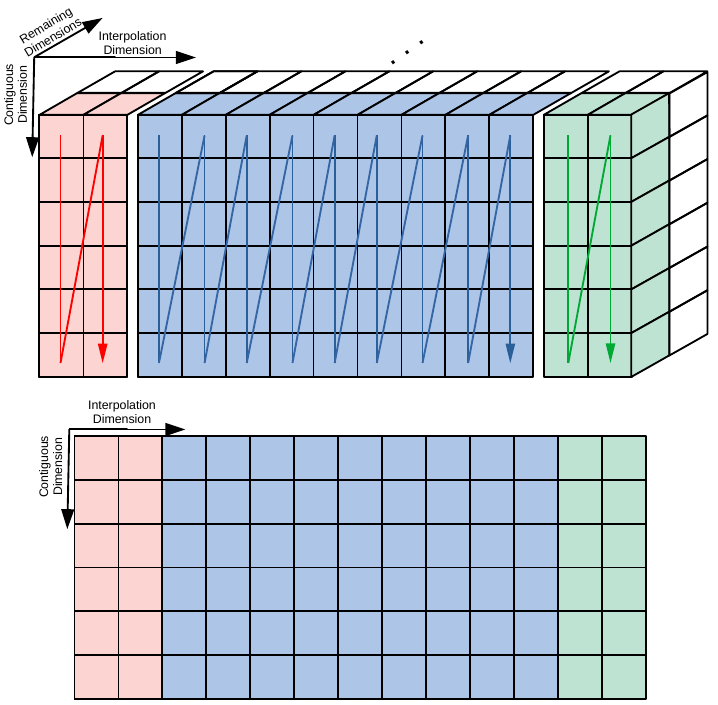}
		\caption{\label{fig:iterationPattern}Illustration of the prefetching needed for the interpolation. From a 6-D View 2-D slices are collected into one continuous 2-D View following the continuous dimension. On GPU this approach can be compared to the usage of shared memory in the matrix transpose example of Nvidia by \citet{MatrixTranspose2022}.}
	\end{center}
\end{figure}

First we need to ensure continuous memory access when applying the Lagrange interpolation stencil. Since the memory of the halo regions and computation domain are not stored continuously in memory it is not possible to iterate with a single set of nested loops. A solution to this problem was already introduced by  \citet{Kormann2019}. A 2-D slice of halo and computation domain is prefetched into continuous memory. Then the interpolation stencil works on the prefetched 2-D slice which resides continuously in memory. The 2-D \verb`View` has to be small enough to fit into high level caches such that every data point is only written and read once. Figure~\ref{fig:iterationPattern} illustrates this approach. The arrow indicates the direction of the iteration. Two dimensions are involved, which are the interpolated dimension and the dimension which is continuously in memory. Therefore, the iteration pattern has to be adapted with every interpolation. Using scratch memory from \citet{Trott2022} provided by \texttt{Kokkos} within the \texttt{TeamPolicy} ensures that the data remains close to the caches.

A second optimization step concerns the amount of \texttt{Kokkos} scratch memory which is allocated. On GPUs the scratch memory space corresponds to shared memory. GPU shared memory is limited to a few kilo-Bytes of explicitly managed cache memory. Using high amounts of shared memory reduces the number of active warps on a GPU. If the shared memory usage increases the number of active warps is reduced and with it the utilization of GPU compute capabilities. With Nsight Compute the shared memory usage has been analyzed and the occupancy of the GPU is limited through shared memory usage. Therefore, the allocation of scratch memory is reduced by writing interpolated grid points directly back into the \verb`View` which holds the computation domain instead of writing the data into a second cache array as in \citep{Kormann2019}.

\subsection{Software architecture of the Lagrange interpolation}
\label{subsec:software_arch_lag_intrp}
In this section a software architecture for the Lagrange interpolation of subsection~\ref{subsec:Lagrange_Interpolation} is presented. Since the Lagrange interpolation is the computationally most intensive calculation of the BSL6D algorithm an efficient implementation is needed. At the same time new physical models which define the interpolation shift $\alpha $ through the characteristic equations~\eqref{equ:characteristic_odes} shall be easy to add. In addition different orders of Lagrange polynomials need to be available for the interpolation. Finally, adding new stencils and characteristics should not require knowing about the iteration pattern from subsection~\ref{subsec:performance_considerations}. In the following we will discuss separate components of Alg.~\ref{alg:LagrIter}, which is a pseudocode of the implemented algorithm.

The generic algorithm has the same structure as the template method which is described in the book of \citet{Gamma2007}. The template method defines a skeleton algorithm which defers primitive operations to separated classes or functions. In case of the interpolation the iteration pattern provides the skeleton algorithm. Primitive operations can be used to abstract $\alpha $ and the stencil. Abstractions of these primitive operations need to share common interfaces.

First the iteration pattern providing the skeleton algorithm is discussed. Lines 10 to 12 and 13 to 15 prefetch 2-D slices from \texttt{SubView}s in scratch memory \texttt{View}s named \verb|local| as is shown in figure~\ref{fig:iterationPattern}. The last two nested loops ranging from 16 to 19 finally carry out the actual interpolation. These iterations work on the index, which is interpolated, and the index, which is continuous in memory. The outer loop iterates on the remaining indexes to complete the interpolation on each point of the phasespace grid.

We now need to define common interfaces for the primitive operations. The first primitive operation calculates the shift $\alpha$ to carry out the interpolation. As discussed in subsection~\ref{subsec:Lagrange_Interpolation} the shift is defined through a generalized interface receiving all six indexes to identify the interpolated grid point and returns the shift $\alpha$ for this grid point. With $\alpha $ the corresponding grid point in the 2-D slice can be updated. Since the \texttt{View} \verb|local| resides continuous in memory we can deduct required neighboring points from the interpolated point. Therefore, the stencil only receives the memory address of the point to be interpolated as well as $\alpha$ and returns the interpolated value. With this we have two interfaces defined for the primitive operations.

\begin{figure}[h!]
\begin{center}
\includegraphics[scale=0.3]{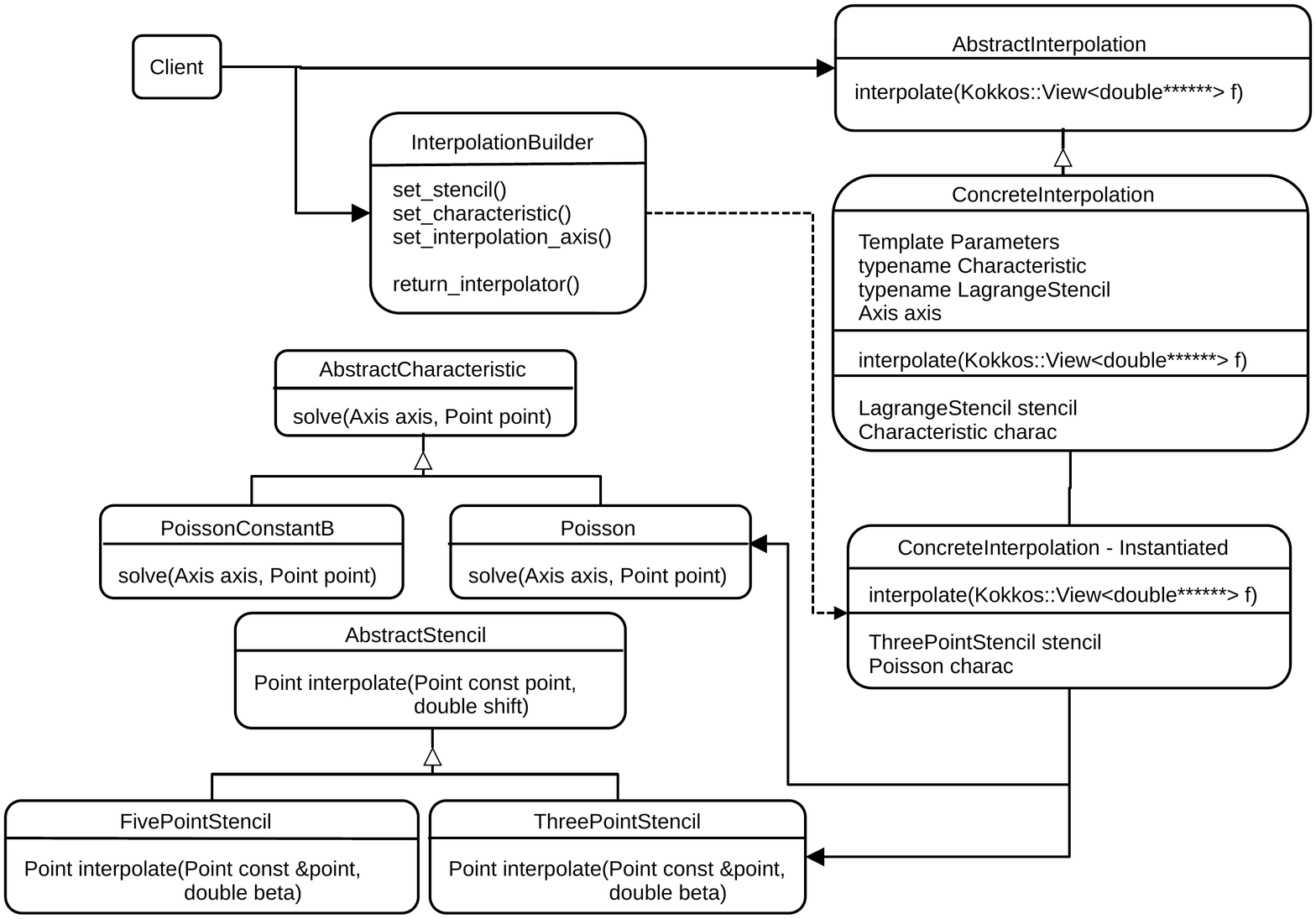}
\caption{\label{fig:classDiagramm}Class diagram of the interpolation using a local stencil to execute the interpolation. All combinations of the concrete interpolations are created at compile time. The interpolation algorithm is shown through pseudocode to point out the template method given in Alg.~\ref{alg:LagrIter}.}
\end{center}
\end{figure}

Now a class structure for Alg.~\ref{alg:LagrIter} is presented. An abstract characteristic class with the virtual function \verb|shift| and an abstract stencil class with the virtual function \verb|interpolate| is declared. All classes which inherit these interfaces can be used within the interpolation class. The interpolation class combines the primitive operations with an interpolated dimension into Alg.~\ref{alg:LagrIter}. The class \verb|Interpolation| therefore interpolates the 6-D distribution function. A concrete interpolation can be created using a builder pattern described by \citet{Gamma2007} through which a client can set all components separately. We omit performance penalties due to abstractions through the usage of static polymorphism. The interpolation class contains \texttt{C++} template parameters for the concrete characteristic class, concrete stencil class and the interpolation dimension. The instantiation of concrete interpolation classes is done within the builder pattern through template meta programming. These implementation details are omitted for simplicity in this paper. The full software architecture is shown in figure~\ref{fig:classDiagramm}.

If new characteristics are added, no changes of the iteration pattern or the stencil are necessary. The same is valid if new stencils are added. Our structure also allows writing separate tests for each component which simplifies testing significantly. The next section evaluates the shared memory performance of the implementation.
\begin{algorithm}[h!]
\KwIn{Distribution function $f(\phasespacevar , 0)$, lower boundary $low(\phasespacevar )$, upper boundary $up(\phasespacevar )$, characteristic, stencil}
\KwOut{Distribution function $f(\phasespacevar , \Delta t)$}
\For{$\{i,j,k,l\}<\{N_{\text{outer},1},N_{\text{outer},2},N_{\text{outer},3},N_{\text{outer},4}\}$}{
local = Kokkos::View($N_{x},N_{z}+4$)\\
sliceLow = get\_ 2D\_ slice($low,i,j,k,l$)\\
sliceF = get\_ 2D\_ slice($f,i,j,k,l$)\\
sliceUp = get\_ 2D\_ slice($up,i,j,k,l$)\\
// boundary width considering \\
// sizeof(sliceLow(0,:)) = sizeof(sliceUp(0,:))\\
bw = sizeof(sliceLow(0,:))\\
\For{$i_1 < \text{bw}$}{
\For{$i_0 < N_\text{continuous}$}{
local($i_0,i_1$) = sliceLow($i_0,i_1$)\\
local($i_0,N_\text{continuous}+\text{bw}+i_1$) = sliceUp($i_0,i_1$)
}
}
\For{$i_1 < N_\text{interpolation}$}{
\For{$i_0 < N_\text{continuous}$}{
local($i_0,\text{bw}+i_1$) = sliceF($i_0,i_1$)
}
}
\For{$i_1 < N_\text{interpolation}$}{
\For{$i_0 < N_\text{continuous}$}{
$\alpha =$ characterisitc.shift$(i_0,i_1,i,j,k,l)$\\
sliceF($i_0,i_1$) = stencil.interpolate(local($i_0,i_1$),$\alpha$)
}
}
}
\caption{\label{alg:LagrIter}1-D Interpolation of the distribution function based on an abstracted shift and interpolation stencil assuming the first dimension of the View is continuous in memory.}
\end{algorithm}

	\section{Shared memory performance measurement}
\label{sec:shared_performance}
\begin{table}[t]
\begin{tiny}
\begin{tabular}{lccc}
\hline
\multicolumn{1}{c}{System}                                                              & COBRA                                                                     & RAVEN                                                                          & AMD System   \\ \hline
CPU                                                                                     & \begin{tabular}[c]{@{}c@{}}2x Intel Xeon Gold 6148\\ Skylake\end{tabular} & \begin{tabular}[c]{@{}c@{}}2x Intel Xeon Platinum 8360Y\\ IceLake\end{tabular} & 2x AMD EPYC 7763                                                  \\
Memory (DDR)                                                                            & 96GB                                                                      & 512 GB                                                                         & 512 GB                                                            \\
Memory Bandwidth - CPU (BabelStream)                                             & 100 (74) GB/s                                                                  & 135 (150) GB/s                                                                       & ---                                                              \\
\begin{tabular}[c]{@{}l@{}}Network Bandwidth uni-directional\\ CPU Nodes (OSU-Benchmark H-H)\end{tabular} & \begin{tabular}[c]{@{}c@{}}OmniPath\\ 12.5 GB/s\end{tabular}              & \begin{tabular}[c]{@{}c@{}}InfiniBand HDR\\ 12.5 (10) GB/s\end{tabular}             & ---                                                              \\
GPU                                                                                     & 2x Nvidia V100                                                            & 4x Nvidia A100                                                                 & 4x AMD MI250                                                      \\
Memory (HBM GPU)                                                                        & 32 GB                                                                     & 40 GB                                                                          & 128 GB                                                            \\
Memory Bandwidth - GPU (BabelStream)                                                                 & 900 (830) GB/s                                                                  & 1550 (1360) GB/s                                                                      & 3200 (2400) TB/s                                                          \\
GPU direct (OSU-Benchmark D-D)                                            & ---                                                                      & 100 (80) GB/s                                                                       & ---                                                             \\
\begin{tabular}[c]{@{}l@{}}Network Bandwidth uni-directional \\ GPU Nodes (OSU-Benchmark D-D) \end{tabular} & \begin{tabular}[c]{@{}c@{}}OmniPath\\ 12.5 (GB/s)\end{tabular}            & \begin{tabular}[c]{@{}c@{}}InfiniBand HDR\\  25 (25) GB/s\end{tabular}              & \begin{tabular}[c]{@{}c@{}} --- \\ --- \end{tabular} \\ \hline
\end{tabular}%
\caption{\label{tab:systems_under_consideration}Specification of the hardware used for performance evaluation. On the AMD system we only considered the GPU capabilities which will therefore be the only given configurations.}
\end{tiny}
\end{table}
This section analyzes the performance of the previously presented software architecture which enabled easy extension and testing of the implementation. The results are obtained by executing the 1-D Lagrange interpolation in all six dimensions of the 6-D distribution function using random shifts $\alpha/\Delta x\in [-1,1]$. The interpolation is executed using uneven stencils with widths three, five, seven and nine. Our target architectures are Nvidia GPUs and Intel CPUs due to the systems on which the production runs are executed. Results on AMD GPUs demonstrate the hardware independent implementation. The systems on which we generated the data are the Cobra GPU and CPU partition and a System containing AMD MI250 GPUs to which we received access. The technical specifications are given in table~\ref{tab:systems_under_consideration}.

The performance of our implementation is measured through hardware counters which are collected using the following tools. On CPU the \texttt{LIKWID} toolkit \cite{Treibig2010} is used, \texttt{NSight Compute} \cite{NsightWeb} for Nvidia GPUs and \texttt{rocprof} \cite{ROCmProf} for AMD GPUs. With all tools we collect walltime, amount of data transferred between main memory and Caches as well as the bandwidth. The \texttt{ROCM} profiler of AMD does not yet contain a derived metric to collect the bandwidth. But the formula used by \texttt{LIKWID} and \texttt{NSight Compute} is used to calculate the bandwidth. Both divide the data read from and written to main memory by the walltime of the kernel.

All calculations are done with 37 points per dimension which results in $37^6$ grid points in the phasespace grid.

\subsection{Walltime measurement of advection kernel}
\begin{figure}
\begin{center}
\includegraphics[scale=0.65]{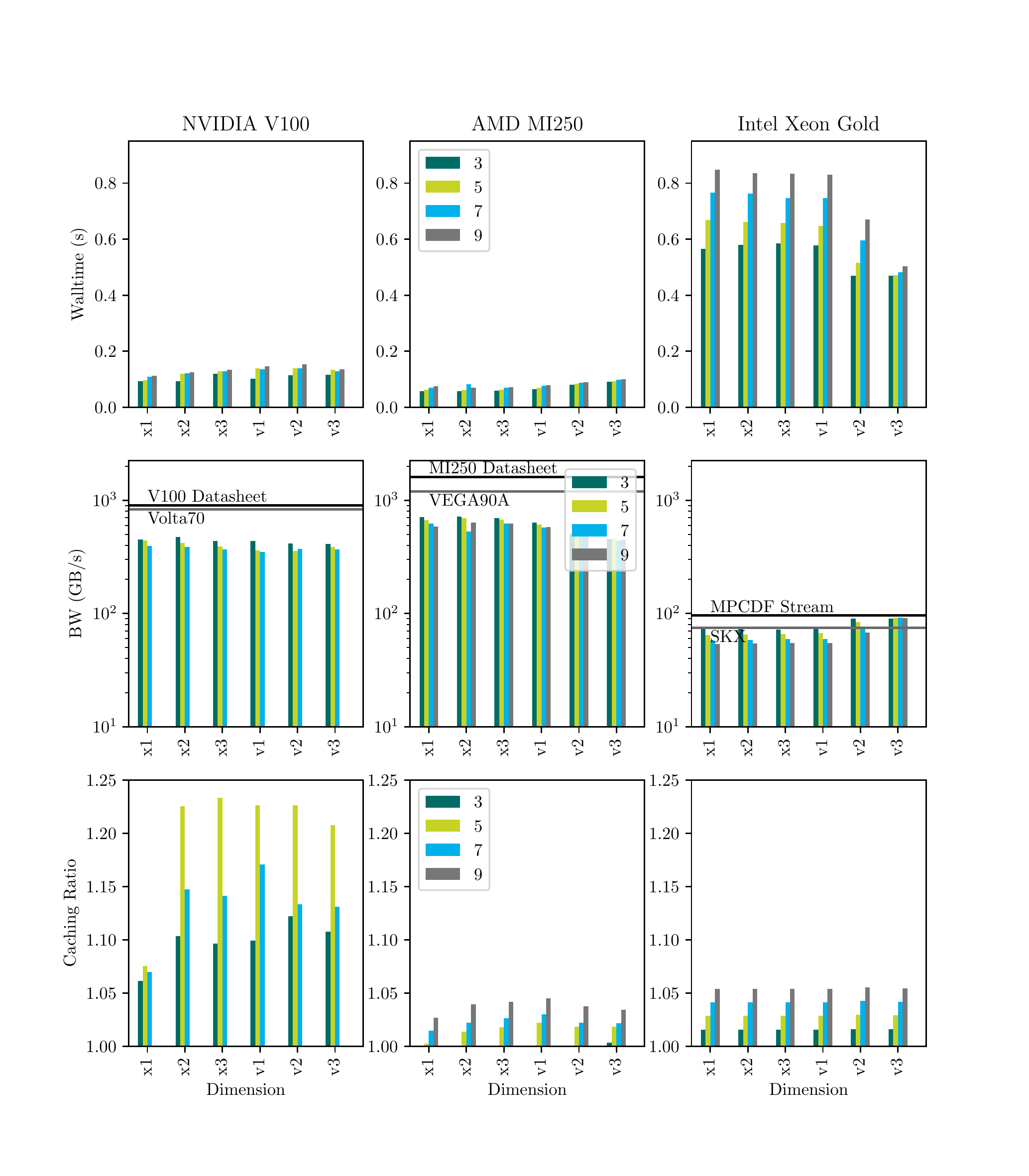}
\caption{\label{plot:sharedMemPerf} Metrics used to evaluate performance portability of the BSL6D implementation. We compare walltime, memory bandwidth and read write data caching ratio for interpolations in all six phase space dimensions. The memory bandwidth is compared to the technical specification for V100 and one core of the MI250 through the black line. The black reference line for the Intel Xeon Gold is compared against a Stream benchmark done by the MPCDF an communicated internally. The grey line corresponds a run using the BabelStream with the \texttt{Kokkos} backend. The \texttt{Kokkos} architecture keyword used to build the BabelStream is used as a label to the grey line.}
\end{center}
\end{figure}

First the walltime behavior of the kernels on the analyzed platforms is compared. The results are plotted in the first row of figure~\ref{plot:sharedMemPerf}. On the ordinate the walltime has been plotted against the dimension in which the interpolation is executed on the abscissa. Two distinct features are observed.

Both GPUs are significantly faster than the CPU due to the high parallelization potential of the BSL6D algorithm. The Nvidia GPUs gain a factor of 3.5 to almost 7.5 while with AMD GPU the walltime is reduced by a factor of 4 to almost 8 compared to the CPU system.

Secondly we can observe an increasing walltime going from dimension $v_3$ to dimension $x_1$ on CPUs while on GPUs this is inverted. Here an increasing walltime is observed going from $x_1$ to $v_3$. This can be explained by the longer strides of the interpolated dimension. On CPUs the contiguous dimension is $v_3$ due to the default memory layout of a \texttt{View}. On GPUs the $x_1$ dimension resides contiguous in memory. The length of strides increases when moving further away from the contiguous dimension.

Comparing the memory bandwidth of the technical specifications of table~\ref{tab:systems_under_consideration} a speedup around nine is achieved if comparing the Intel Gold CPU to the Nvidia V100 GPU specifications and a factor of 16 compared to half a core of the AMD MI250 GPU. Further optimization is therefore necessary to utilize the full speedup potential of the GPU. Further insights into explicit performance metrics are provided in the next subsection.

\subsection{Performance insights in the advection kernel}
In this subsection a look into the memory bandwidth and the caching behavior is taken. The memory bandwidth is calculated by dividing the data transferred between main memory and caches by the walltime of the kernel. The caching is analyzed by calculating the expected amount of transferred data to the measured data transfer of hardware counters. In case of optimal caching the ratio of both values should be one. For the Nvidia GPU we had to skip the stencil of width nine, since the memory requirements exceeded the available memory when collecting data with \texttt{NSight Compute}.

The bandwidth of the three different systems are given in the second row of figure~\ref{plot:sharedMemPerf}. The plot has a similar structure as the walltime plot used in the previous subsection. The Bandwidth (BW) is plotted on the ordinate against the interpolated dimension on the abscissa. The horizontal black line marks the peak bandwidth given in the technical specifications for both GPUs. In the CPU plot the black line has been measured by the MPCDF on the Cobra cluster and was communicated internally. For Intel CPUs no bandwidth specifications of the vendor could be found. The gray line has been obtained through a simple Stream benchmark using BabelStream developed by \citet{Deakin2018} build with the \texttt{Kokkos} backend as a reference for the achievable bandwidth. From the results of the BabelStream benchmark we took the highest bandwidth which has been achieved with the Triad Stream on all Systems. The AMD MI250 contains basically two separate GPUs on a single device. Since in this example only a single MPI process is used for AMD, the data is compared to half the bandwidth given in the technical specification.

On both GPU systems about a factor of two is missing to the peak bandwidth. The BabelStream Triad Benchmark achieves more than 90\% of the manufacturer specifications. The best performance results of the observed systems have been achieved for the CPU system. Two observation have been made considering the GPU performance. The scratch memory, which has already been discussed in subsection~\ref{subsec:performance_considerations}, can reduce performance if large amounts of scratch memory are allocated. In addition, \texttt{View}s and \texttt{ExecutionPolicies} have a high registers consumption on the GPU. Register usage of \texttt{View}s could be explained through integer arithmetic which is necessary to map multiple indexes to a specific memory address. Therefore, scratch memory and register usage reduce the occupancy on GPU while integer calculations require additional instructions cycles on the GPU.
%This is due to the fact that the implementation follows \citep{Kormann2019} which has been optimized for CPU usage. \kk{Is this really true. Are the optimizations in that paper not equally relevant on GPUs but it is simply even harder on GPUs to get to the optimal performance? If not, which would be the steps that would be needed for optimization on GPU?} In the contiguous dimension we achieve up to 90\% of the peak bandwidth and are thus even above the bandwidth achieved with the BabelStream compiled with \texttt{Kokkos}.

Lastly a look into the caching ratio of the implementation in the third row of figure~\ref{plot:sharedMemPerf} is taken. The caching ratio is given as the amount of data transferred between caches and main memory divided by the expected amount of transferred data if every value of the distribution function is only read from and written to memory once. The data on AMD GPUs might not be fully reliable since for a stencil with width three the caching ratio smaller than one has been encountered which is not possible. On CPUs we almost achieve optimal caching ratio. Less than $5\%$ of the data is read multiple times. The caching on Nvidia GPUs does not work as well. Still more than $75\%$ of the data is cached properly and is read only once.
	\section{Distributed memory performance measurement}
\label{sec:distributed_performance}
Due to large memory requirement of the 6-D problems to run high resolution simulations it is necessary to make use of multiple nodes. The interesting metric to increase resolutions of our grid is weak scaling. %Since the main bottleneck is halo communication in our application, it is sensible to use a large amount of the available memory on each process in order to keep the ratio of halo cells as small as possible.

In the following we will investigate the walltime behavior of the BSL6D code when running simulations on multiple nodes. We perform the scaling examples on the two node level architectures of the Raven cluster described in table~\ref{tab:systems_under_consideration}. These two node level architectures allows to compare the walltime behavior on an accelerator based system containing Nvidia A100 GPUs connected using GPU direct technology and a CPU based system containing Intel IceLake Processors.

\begin{table}[]
\begin{center}
\begin{tabular}{ccc|cccccc}
\multicolumn{1}{l}{\# MPI} & \multicolumn{1}{l}{\# CPUs} & \multicolumn{1}{l|}{\# GPUs} & \multicolumn{6}{l}{MPI per Dim x1 to v3} \\ \hline
1                          & 18                          & 1                            & 1     & 1    & 1    & 1    & 1    & 1    \\
2                          & 36                          & 2                            & 2     & 1    & 1    & 1    & 1    & 1    \\
4                          & 72                          & 4                            & 2     & 2    & 1    & 1    & 1    & 1    \\
8                          & 144                         & 8                            & 2     & 2    & 2    & 1    & 1    & 1    \\
16                         & 288                         & 16                           & 2     & 2    & 2    & 2    & 1    & 1    \\
32                         & 576                         & 32                           & 2     & 2    & 2    & 2    & 2    & 1    \\
64                         & 1152                        & 64                           & 2     & 2    & 2    & 2    & 2    & 2   
\end{tabular}
\caption{\label{tab:setup_weak_scaling} Setup used for the weak scaling with $32^6$ grid points per dimension.}
\end{center}
\end{table}
The simulation is set up in the following way: In all six dimensions we will use $N=32$ points per MPI process. For the weak scaling we start with a single MPI process and no distributed dimension. The number of MPI processes is doubled by parallelizing one more dimension with every step. The weak scaling setup is shown in table~\ref{tab:setup_weak_scaling}. The distance of two MPI processes in the raises from $v_3$ to the $x_1$ dimension. Therefore, in the two highest dimensions which are run parallel we can use GPU direct communication.

For all simulations we use a stencil with seven points such that three points are contained in each of the halo regions of the interpolated dimension. The ratio of computation to communication time will be analyzed in the subsection~\ref{subsec:comm_mem_bottleneck}.

\subsection{Scaling}
\label{subsec:weak_scaling}
\begin{figure}
\includegraphics[scale=.7]{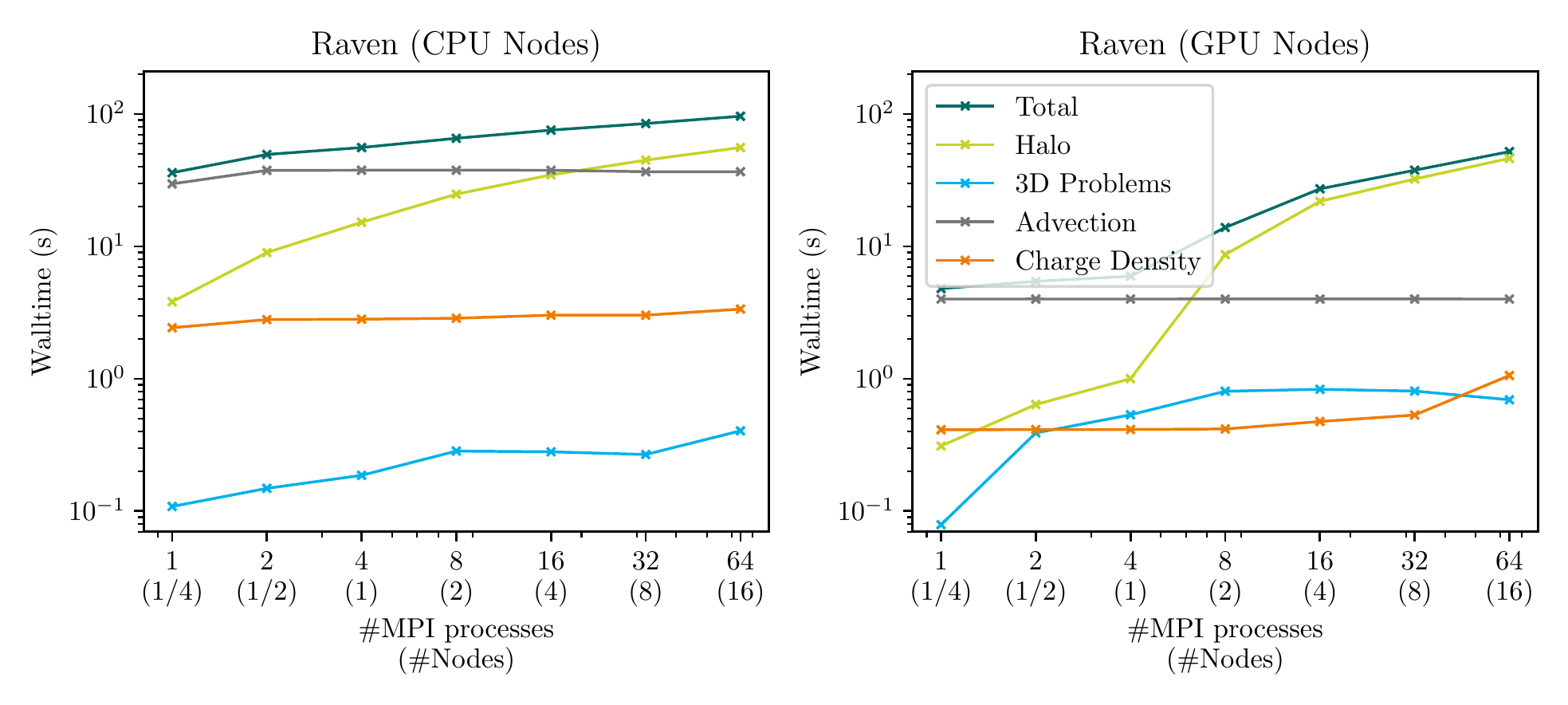}
\caption{\label{plot:scaling}Scaling of BSL6D Code to multiple CPU nodes in the left plot and GPU nodes in the right plot.}
\end{figure}

We first consider the scaling of our code to multiple CPU and GPU nodes. The scaling results of the previously described experiment is given in the left plot of figure~\ref{plot:scaling}. During the weak scaling we observe the behavior of the four categories defined in the subsection~\ref{subsec:bsl6d_algorithm}. Both plots show the walltime of advection, halo communication, solving the 3-D problems and the reduction of the distribution function into the density. The walltime of the advection stays constant during all runs. This is to be expected, since the problem size per MPI process stays the same during the weak scaling. Increasing the parallelism naturally increases the cost for components including MPI communication. Three different types of MPI interactions can be found and are given in figure~\ref{fig:flowchart}. First the halo communication is a point to point communication. Secondly the reduction of the distribution function into the density involves a reduction on all distributed dimensions of $\vvec $. Solving all 3-D problems is based on the FFT algorithm. This requires an all to all MPI communication on all dimensions of $\xvec $ which are distributed. The latter two types which solve for $\rho$ and $\vecb{E}$ are negligible throughout all steps compared to the halo communication and advection when comparing against the results of figure~\ref{plot:scaling}. In the following we will therefore focus on the halo communication. Two sections are found in which the walltime for halo communication increases linearly with a fixed slope. The first section to be identified is going from one to four MPI processes where only intranode communication between different MPI processes is necessary. On GPU the slope is still small due to GPU direct communication. The second section starts when moving from intranode to internode communication starting with eight MPI processes. Here the slope becomes steeper, since the internode bandwidth is small compared to the intranode bandwidth. Specifically when moving from GPU direct technology to internode communication the reduction of bandwidth results in a significant increase of time needed for the halo communication.

Especially in the GPU case we can observe that the runtime is dominated by communication. But also for CPU implementation the communication time supersedes the computation of the interpolation if more than two dimensions have to execute internode communications. The communication bottleneck will be quantified in the next subsection.

\subsection{Comparing network to computation hardware specifications}
\label{subsec:comm_mem_bottleneck}
In the previous subsection we observed that the communication supersedes computing time of the advection already for a few nodes. Therefore, we take a look into the hardware limitations of MPI communication. In this subsection the computing performance of the processing unit is compared to the network capacity.

The local update formula of the advection can be compared to a stencil algorithm as discussed in subsection~\ref{subsec:Lagrange_Interpolation} which is generally memory bound. In the roofline model developed by \citet{Williams2009} memory bound algorithms are limited through the exchange of data between main memory and the processing unit. An algorithm is memory bound if the number of floating point operations executed per byte taken from memory also defined as operational intensity are below the machine balance. Machine balance is the arithmetic intensity at which a memory bound algorithm becomes compute bound.

If we now introduce a distributed memory concept using MPI, another bottleneck has to be taken into account which is not defined in the Roofline model. The walltime of the advection can be limited by the halo communication if exchanging data of halo regions takes longer than applying the stencil to all degrees of freedom. We compare the memory bound and communication bound bottlenecks by calculating the theoretically minimal achievable runtime for both components. In the following we give two expressions to calculate the theoretical runtime by dividing the transferred amount of data by the bandwidth.

For this we consider again a hypercube with $N$ points per dimension and a width $w$ of the halo region. The minimal walltime needed for the execution of the interpolation is given by
\begin{align}
	\label{equ:walltime_mem}
	t_{\text{interpolate}} = \dfrac{\left(N^6 \cdot (N_\text{read} + N_\text{write}) + N^5 \cdot w \cdot N_\text{read} \cdot N_{halo} \right) \cdot \SI{8}{B}}{\text{BW}_\text{PU}}
\end{align}
where $\text{BW}_\text{PU}$ is the memory bandwidth of the processing unit in $[\text{BW}_\text{PU}] = \si{B\per s}$, $N_\text{read}$ and $N_\text{write}$ defines how often the distribution function is read from and written to memory during the interpolation. The first term in the numerator of equation~\eqref{equ:walltime_mem} considers the memory transfer of the distribution function while the second term considers the halo regions. $N_{\text{halo}}$ separate halos are needed to calculate the interpolant. A similar formula can be given for the halo communication through MPI
\begin{align}
	\label{equ:walltime_comm}
	t_\text{comm} = \dfrac{N^5 \cdot w \cdot (N_\text{read} + N_\text{write} ) \cdot N_{halo} \cdot \SI{8}{B}}{(BW_\text{Network}/N_\text{MPI})}
\end{align}
where we consider the same setup as for $t_\text{interpolate}$ with the difference that the bandwidth of the Node $\text{BW}_\text{Network}$ has to be shared between the MPI processes pinned to this Node. The maximum of these two theoretical walltimes limits the walltime of the implementation.

\begin{figure}
	\begin{center}
		\includegraphics[scale=0.7]{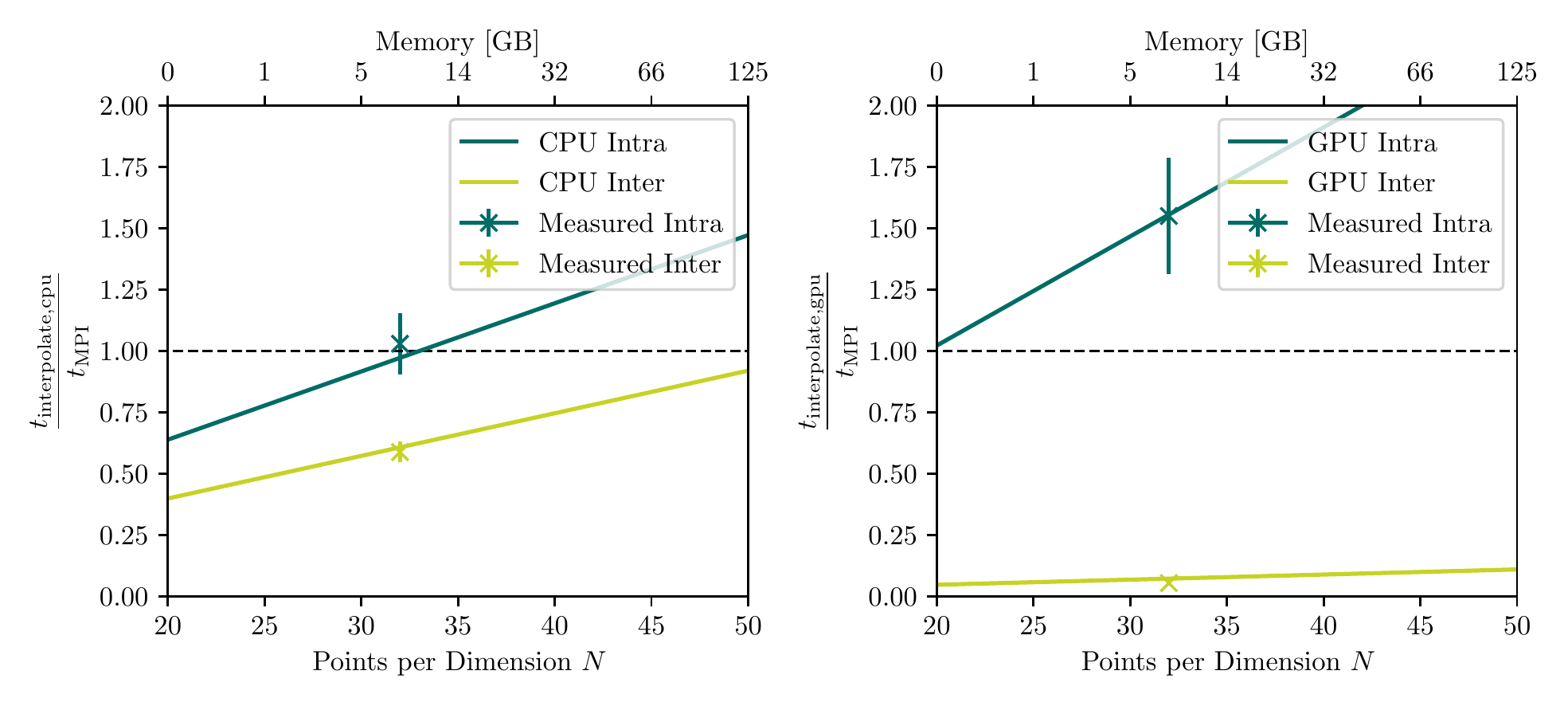}
		\caption{\label{plot:commMemBound}Communication to memory bound code execution. Below one the implementation is network communication bound and above one it is bound by computation. The upper abscissa gives the memory consumption needed for the 6-D distribution function. We consider $N_\text{read}=N_\text{write}=1$ and $N_\text{halo}=2$ with a stencil of width seven requiring a boundary of width $w=3$. The bandwidth available is given in table~\ref{tab:systems_under_consideration}. The plot shows computing to communication time ratio considering intra node and internode MPI communication based on equations~\eqref{equ:walltime_mem} and \eqref{equ:walltime_comm}}.
	\end{center}
\end{figure}

In figure~\ref{plot:commMemBound} the ratio of equation~\eqref{equ:walltime_mem} to \eqref{equ:walltime_comm} is plotted for the straight lines. Values above one are bound through the interpolation while below one the problem is bound through network communication. The lines are calculated for the setup of the weak scaling above using a stencil of width 7 with $w=3$, $N=32$ and the OSU-Benchmark specifications of table~\ref{tab:systems_under_consideration} for the Raven cluster. The value $\text{BW}_{PU}$ has been measured on Raven for the A100 as in section~\ref{sec:shared_performance} with the same result, that we achieve 90\% of the CPU peak bandwidth and about 50\% of the GPU peak bandwidth. The algorithm is bound through the interpolation only in case of intranode communication. As soon as internode communication is required, the advection is strongly bound by the network communication bottleneck.

This allows us to compare the model to our scaling results. We calculated average and standard deviation of the time ratio $t_\text{interpolate}/t_\text{MPI}$ for the scaling with $N=32$ in figure~\ref{plot:scaling} and add the corresponding point to the plot in figure~\ref{plot:commMemBound}. We can observe that the previously  presented model for the communication bottleneck matches our measured data quite well. This result shows that the algorithm does not scale properly to large compute clusters since the scaling efficiency will decrease significantly due to the network communication bottleneck.

	\section{Physical verification through Ion Bernstein waves}
\label{sec:physical_validation}
In this section, we verify the implementation by reproducing the dispersion relation $\omega(k)$ of ion Bernstein waves found by \citet{bernstein58}. As ansatz, a plane wave
\begin{align}
	 \phi(\xvec, t) = \hat\phi_{\vec k} \exp( i (\vecb{k}\cdot\xvec - \omega t))
\end{align}
is used to find the solutions in ($\omega, \vecb{k}$) for the following Vlasov model.

The simulation is based on the same setup as the kinetic simulations of \citet{sturdevant18}. For this test simplified field equations are used. Instead of simulating electrons with static ions in the background, ions ($q = 1$) are simulated with adiabatic electrons, simplifying the field equations~\eqref{equ:poisson} for the electrostatic potential to
\begin{align}
	\dfrac{n_0q^2}{T_e}\phi(\xvec , t) = \rho(\xvec, t),
\end{align}
where $T_e$ denotes the temperature of the adiabatic electrons. The distribution function is initialized using a product approach based on white noise for the configuration space and a Maxwellian distribution in velocity space
\begin{align}
	f(\phasespacevar , t=0) = f_{x,0}(\xvec ) f_{v,0}(\vvec ) = n_0(1+\text{rand}(\xvec )) \left(\dfrac{m}{2\pi T_i}\right)^{3/2}\exp\left( -\dfrac{m \vvec^2}{2T_i}\right)
\end{align}
where $m$ corresponds to the ion mass and $T$ to the temperature of the system. The magnetic background field is chosen as $\vecb{B}_0 = B_0 \hat{\vecb{e}}_z$. The calculations are simplified by setting all physical constants to one $(q=m=T_{i}=T_e=B_0=n_0=c=1)$. The function $\text{rand}(\xvec)$ creates white noise with a small amplitude $\epsilon\ll 1$.

Ion Bernstein waves (IBW) have a real frequency and are not subject to growth or
damping. We can extract them by simply run the simulation until all other waves 
in the system are damped, leaving only the IBWs. It suffices to study only wave 
vectors $k_x$ with $k_y=k_z=0$, reducing the dimensionality of the problem. We 
nevertheless require a few grid points in $y$ and $z$ dimensions to accomodate 
the minimal stencil widths. Therefore, the system is set up using a
$128\times 8^2\times 32\times 16^2$ grid, a configuration space box length
$L_\xvec = \frac{20}{3}\pi \times 2\pi \times 2\pi $ and a velocity space with
maximum velocity $v_{\text{max}}= 4 v_{\mathrm{th}}$ for all directions. The
time step is chosen as $\Delta t = 0.025$.

\begin{figure}
\begin{center}
\includegraphics[scale=0.6]{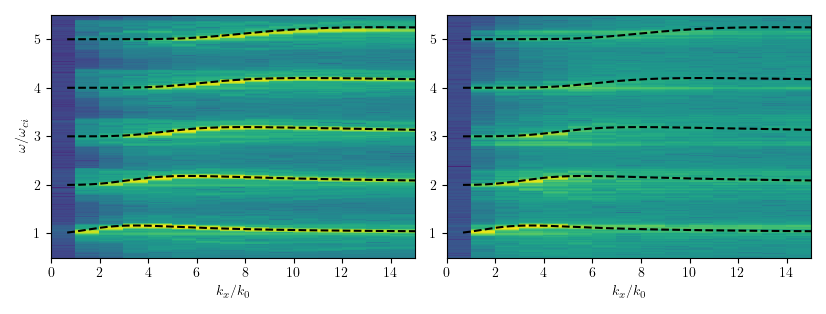}
\end{center}
\caption{\label{fig:ion_bernstein_wave_dispersion} Frequency-wave number
  spectrum of electrostatic potential after relaxation, exhibiting the ion
  Bernstein wave dispersion relation for the interpolation stencil width
  $d = 8$ (left) and $d=4$ (right) (color coded) and analytical dispersion
  relation for the IBWs (dashed).}
\end{figure}

After the simulation has run sufficiently long the dispersion relation of IBWs
can be extracted by Fourier transforming the density perturbation in time and
space. Figure~\ref{fig:ion_bernstein_wave_dispersion} shows the dispersion
relation that has been obtained by transforming the time interval from
$t_1=4000$ to $t_2=5000$. The plot displays clear branches of the dispersion
relation $\omega(k_x)$ close to every harmonic of the Larmor frequency
$\omega_{{ci}}$. The frequencies of the IBWs are slightly shifted upwards
compared to corresponding harmonics of the Larmor frequencies
$\omega_{{ci},m} = m\omega_{ci} = m$ and converge to the harmonics for large
wavenumbers,
\begin{align}
\lim_{k_x \rightarrow\infty} \omega(k_x) = \omega_{ci,m} = m.
\end{align}
The setup was run twice using different interpolation stencils to show the
different diffusive behavior of the numerical scheme. Using high order
stencils is important to resolve multiple modes. Almost three times the number
of modes can be observed by increasing the interpolation order from 3 to 7.
The reason for this is the strong diffusive behavior of local interpolation
schemes. Since higher order stencils exacerbate the communication bottleneck
described previously, a compromise case has to be found between high order 
stencils and resolution of modes.

The dispersion relation for this system can be derived following the
computation in the book of \citet{hazeltine2018}. In the limit of
$k_z= \vec k \cdot \hat{\vec z} =0$ the dispersion relation is given by
\begin{align}
	- \omega\sum_{m \in \mathbb Z} \frac{e^{-k_x^2}I_m(k_x^2)}{\omega + m}-2 = 0
\end{align}
where the $I_m(x)$ denote the modified Bessel functions. The dispersion
relation for $\omega( k_x )$ has one solution for every term $m$, close to
$\omega_{{ci},m}$. In figure~\ref{fig:ion_bernstein_wave_dispersion} the analytical
dispersion relations (dashed lines) are plotted alongside the numerical
results. We can see a clear agreement between the two results.

Reproducing the dispersion relation of ion Bernstein waves makes the
correctness of this model in the high frequency regime of plasma physics close
to the Larmor frequency plausible. These results verify the model for 
phenomena with frequencies close to the Larmor frequency.

	\section{Conclusion}\label{sec:conclusions}
	
	In this paper the implementation of a performance portable semi-Lagrange code for the fully kinetic Vlasov equation has been presented. First, the memory requirement challenge of 6-D problems and the corresponding memory optimizations for our algorithmic setup have been discussed. We implemented the Lagrange interpolation kernel using the 'Template Method' design pattern which leads to a modular class structure. The separate components that have been identified are the iteration, interpolation shift and interpolation stencil. In the main kernel these three components interact through the definition of general interfaces. The modularity allows easy testing and extension of separate components in the algorithm.
	
	Our software architecture achieves good performance throughout several shared memory concepts and hardware architectures due to the usage of the performance portability framework \texttt{Kokkos}. All analyzed architectures can be used while still maintaining a single code base. The implementation leverages the parallelization potential on GPU accelerated nodes which can reduce the runtime of the main kernel by a factor of three to eight compared to CPU nodes. In our node level performance analysis we still identify encounter further optimization potential on GPU architectures. Reducing integer calculations as well as scratch memory and register usage could potentially lead to higher hardware utilization on GPUs. 
	
	Scaling results have been provided and the communication bottleneck of 6D implementations with distributed memory concepts has been quantified, which remains as a challenge. The scaling experiments on both investigated node level architectures prove that the tasks of different MPI processes are not independent enough to scale properly on large compute clusters. We proved through an analysis of the network bandwidth, that this is not solvable through code optimizations since it is a hardware limit. Therefore, different algorithms have to be investigated to further decouple MPI processes and achieve better scaling results. Finally, the implementation has been verified by reproducing ion Bernstein waves. Future work will also target an extension of the physical model by more accurate field equations and geometry, as well as collisions.
	
	\section{Acknowledgements}
	Computations were performed on the HPC systems Raven and Cobra at the Max Planck Computing and Data Facility as well as Marconi100 at Cineca through the LoGy project. In addition, we thank Klaus Reuter for fruitful discussions on the network performance of Raven.
	
	This work has been carried out partly within the framework of the EUROfusion Consortium, funded by the European Union via the Euratom Research and Training Program (Grant Agreement No 101052200 – EUROfusion). Support has also been received by the EUROfusion High Performance Computer (Marconi-Fusion). Views and opinions expressed are however those of the author(s) only and do not necessarily reflect those of the European Union or the European Commission. Neither the European Union nor the European Commission can be held responsible for them.	
	
	\bibliographystyle{elsarticle-num-names}
	\bibliography{./Bibliographie/Bibliographie.bib}

\end{document}